\documentclass[twoside,fleqn]{article}
\usepackage{espcrc2,texdraw,epsfig}
\title{Spectrum of $k$-string tensions in SU(N) gauge theories
}
\author{L. Del Debbio$^a$, H. Panagopoulos$^b$, P. Rossi$^a$, E. Vicari$^a$
\vskip 5mm
$^a$ Dipartimento di Fisica, Universit\`a di Pisa,
and INFN, I-56127 Pisa, Italy \\
$^b$ Department of Physics, University of Cyprus,
                      Nicosia CY-1678, Cyprus}

\begin{document}

\begin{abstract}
We compute, for the four-dimensional 
SU(4) and SU(6) gauge theories formulated on a lattice, the string tensions 
$\sigma_k$ related to sources with $Z_N$ charge $k$, using Monte Carlo simulations.
Our results are compatible with  
$\sigma_k \propto \sin k \pi/N$, and show sizeable
deviations from Casimir scaling.
\end{abstract}

\maketitle

\section{Introduction}

Nonabelian gauge theories are the building blocks of the
field-theoretical description of fundamental interactions. It is
therefore essential to achieve a deep understanding of their physical
content and a better quantitative knowledge of their testable
predictions.  It is widely believed that nonabelian gauge theories
admit a reinterpretation in terms of effective strings; describing
their properties is an issue of the utmost importance. In particular
one would like to understand if these strings belong to some wider
group of objects (``universality class'') whose properties may
eventually be studied by different approaches and techniques.  The
study of the string tensions between static charges in representations
higher than the fundamental one and for different values of $N$ may
shed new light on the nature of the confining strings, helping to
identify the most appropriate models of the QCD vacuum and to select
among the various confinement hypotheses. A static source 
carrying charge $k$ with
respect to the center $Z_N$ is confined by a $k$-string with string
tension $\sigma_k$ ($\sigma_1\equiv \sigma$ is the string tension
related to the fundamental representation).  
The $k$ string is the lightest state propagating in the $k$-charged
channel, and is related to the antisymmetric representation of
rank $k$.
If $\sigma_k < k\,
\sigma$, then a string with charge $k$ is stable against decay to $k$
strings of charge one.  Charge conjugation implies
$\sigma_k=\sigma_{N-k}$. Therefore $SU(3)$ has only one independent
string tension determining the large distance behavior of the
potential for $k\ne 0$.  One must consider larger values of $N$ to
look for distinct $k$-strings.  In particular for $N\geq 4$ one may
consider the ratio
\begin{equation}
R(k,N)\equiv {\sigma_k/\sigma}.
\end{equation}

\section{Models and their predictions}

Some
different conjectures on the behavior of $R(k,N)$ have been discussed
in the recent literature. We briefly discuss a few of them before
presenting the results of our numerical simulations.

\subsection{Casimir scaling}
According to this hypothesis (see Refs.~\cite{AOP-84,Bali-00,deldar00,SS-00,LT-01}):
\begin{equation}
R(k,N) = C(k,N)\equiv \frac{k (N-k)}{(N-1)}
\end{equation}
This formula is exact in two-dimensional SU($N$) gauge theories.
In four dimensions it is satisfied  by
the strong-coupling limit of the lattice Hamiltonian formulation
of SU($N$) gauge theories, and 
by the small-distance behavior of the potential between two static 
charges in different representations, as shown by perturbation theory up to two
loops.

The main objections to Casimir scaling come from the absence of a mechanism 
preserving Casimir scaling from small distance 
(essentially perturbative, characterized by a 
Coulombic potential) to large distance
(characterized by a string tension for sources carrying $Z_N$
charge). Moreover, Casimir scaling
does not survive the next-to-leading order calculation of
the ratios $R(k,N)$ in the strong-coupling lattice Hamiltonian
approach \cite{ldd01}.

\subsection{Sine formula}
Another interesting hypothesis is that
the $k$-string  ratios $R(k,N)$ may reveal a universal behavior within a
large class of asymptotically free theories characterised by 
the $SU(N)$ symmetry \cite{Strassler-98}.
Accordingly, the $k$-string ratios should be 
\begin{equation}
R(k,N) = S(k,N)\equiv\frac{\sin ({\pi k/N})}{\sin({\pi/N})}.
\label{sinf}
\end{equation}
Indeed, this result is obtained
for  the ${\cal N} = 2$ supersymmetric
SU($N$) gauge theory softly broken  to ${\cal N} = 1$
\cite{DS-95,HSZ-98}. 
The same result has been derived in the context of M-theory,
and extended to the case of large breaking of the ${\cal N}=2$
supersymmetric theory \cite{HSZ-98}. 
Moreover, it is suggested by a (rather speculative) M-theory
approach to QCD.

The same formula e\-mer\-ges for the spectrum of the bound states in
the two-di\-men\-sio\-nal $SU(N)\times SU(N)$ chi\-ral models, which
are matrix-valued, asymptotically free, and present interesting
analogies with the four-dimensional gauge theories (see
e.g. Refs.~\cite{Polyakov-88,PCV-98}).  For these models the spectrum
is obtained from the exact S-matrix, derived using essentially the
Bethe Ansatz.

The main objection to this proposal is
essentially the weakness of the
hypotheses on which the universality assumption is based.

\section{Results from Monte Carlo simulations}
We performed numerical Monte-Carlo simulations of four-dimensional
lattice $SU(4)$ and $SU(6)$ gauge theories using the Wilson
formulation. 
Our results were obtained from very high statistics runs for $SU(4)$
(2-4$\times 10^6$ sweeps on $12^3\times 24$ and $16^3\times 32$
lattices).  The statistics
for $SU(6)$ was approximately 10 times smaller. The reader
is referred to Ref.~\cite{ldd01} and a forthcoming paper for the details of the
analysis and for comparison with related work~\cite{LT-01}.
\begin{figure}[tp]
\begin{center}
\epsfig{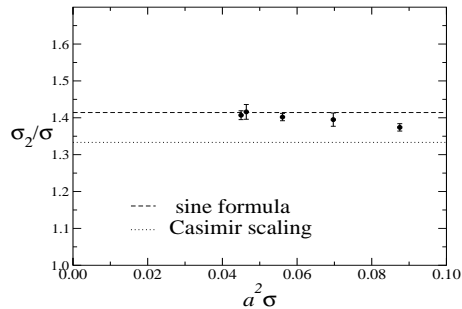} 
\end{center}
\vspace{-1cm}
\caption{$k$-string ratio $\sigma_2/\sigma$ for $SU(4)$.} 
\label{fig:su4}
\vspace{-.8cm}
\end{figure}

In order to compute the $k$-string tensions, we consider
correlators of strings in the appropriate representations:
\begin{equation}
F_k(t) = \sum_{x_1,x_2} \langle\chi_k[P(0,0;0)] \chi_k[P(x_1,x_2;t)]\rangle
\end{equation}
where 
\begin{eqnarray}
&&P(x_1,x_2;t) = \Pi_{x_3} U_3(x_1,x_2,x_3;t) \\
&&\chi_2[P]    = {\mathrm Tr} P^2 - ({\mathrm Tr} P)^2 \\
&&\chi_3[P]    = 2 {\mathrm Tr} P^3 - 3 {\mathrm Tr} P^2 {\mathrm Tr}
P + ({\mathrm Tr} P)^3
\end{eqnarray}
We use standard smearing techniques to improve the overlap with the
lightest propagating state.
The $k$-string tensions are determined from the 
asymptotic decay of the correlators, that, for a $k$-loop of size $L$, 
is~\cite{DSST-85,LT-01}:
\begin{equation}
F_k(t) \sim {\rm exp}-\left(\sigma_k L - {\pi \over 3 L}\right)t,
\end{equation}
where the $O(1/L)$ correction is conjectured to be universal and is
related to the flux excitations described by a free bosonic string
\cite{LSW-80}.  
The choice of the fit-range is a delicate matter: correlations at
short time distances are affected by heavier state contributions,
while at long time distances the signal is obscured by the statistical
noise. A systematic error related to the choice of the fit-range is
therefore unavoidable. To keep it under control, we
perfomed a careful analysis, especially in the case of SU(4), where
the high statistic of the simulations provided good estimates of the correlators
up to relatively large distances.

\begin{figure}[tp]
\begin{center}
\epsfig{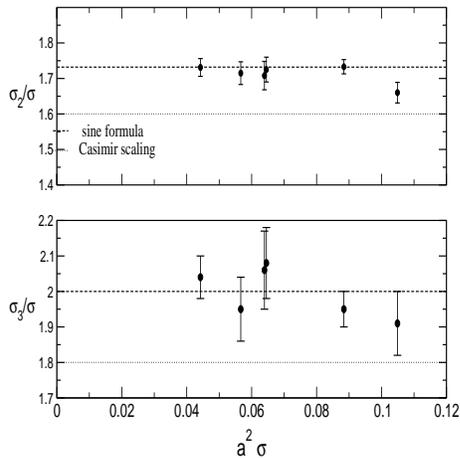} 
\end{center}
\vspace{-1cm}
\caption{$k$-string  ratios for $SU(6)$.} 
\label{fig:su6}
\vspace{-.8cm}
\end{figure}

Results for $R(k,N)$ are shown for $N=4,6$ in Figs.~\ref{fig:su4}
and~\ref{fig:su6} respectively, and plotted
versus $\sigma$.
The tension ratios show a
satisfactory scaling behavior for the coupling values chosen for the
simulation. Therefore we did not find necessary to fit the dependence
of our result on the lattice spacing. 
Our estimates are essentially obtained from the results
at the largest $\beta$-values (smallest $\sigma$ values),
see Ref.~\cite{ldd01}. 
The size of scaling violations can be inferred from the
data at lower values of the coupling; the indication is that they are
comparable with the error quoted below.

Our results for the ratios are
(the SU(4) estimate is still preliminary):
\begin{eqnarray}
R(2,4) &=& 1.405 \pm 0.015\\
R(2,6) &=& 1.72  \pm 0.03 \\
R(3,6) &=& 1.99  \pm 0.07 
\end{eqnarray}
We mention the result $R(2,4)=1.357(29)$ reported in
Ref.~\cite{LT-01},
which is marginally consistent with ours.
We have also explored correlators in the {\it symmetric} rank-2
representation, finding no evidence for stable bound states, 
as expected.

\begin{figure}[tp]
\begin{center}
\epsfig{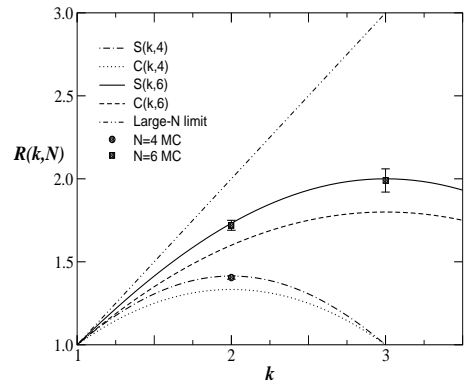} 
\end{center}
\vspace{-1cm}
\caption{
Comparison of the various hypotheses for the $k$-string
ratios with our MC results.} 
\label{summary}
\vspace{-.8cm}
\end{figure}

Figure~\ref{summary} compares our MC results
with the above-mentioned hypotheses of spectrum.  We claim that
$SU(4)$ and $SU(6)$ results show substantial agreement with the sine
formula and evidence of disagreement with Casimir scaling.  Indeed the
sine formula (\ref{sinf}) predicts $S(2,4)=\sqrt{2}=1.414...$, $S(2,6)
= 1.732...$, and $S(3,6)=2$ respectively, while the Casimir scaling
predictions are $C(2,4)=4/3$, $C(2,6)=8/5$ and $C(3,6)=9/5$.
Considering our results alltogether, we can state that the sine
formula is verified within an accuracy of approximately 1\%.  
This result should be relevant for the recent debate on confinement models.
Of course our numerical results do not  prove that
the sine formula holds exactly. But they put a stringent
bound on the size of the possible corrections.
On the other hand, our results appear rather conclusive on 
the existence of deviations from the Casimir scaling.
However, Casimir scaling  may still be considered as a reasonable
approximation, since the largest deviations we observed
were about 10\%.

\end{document}